\documentclass[12pt]{article}

\usepackage{amsmath}
\usepackage{amssymb}
\usepackage{cite}
\usepackage{theorem}
\usepackage{graphicx}
\usepackage{enumerate}
\usepackage{hsumacros}
\usepackage{color}

\theoremstyle{definition}
\newtheorem{definition}[theorem]{Definition}

\definecolor{red}{rgb}{1, 0, 0}

\begin{document}

\title{Iterative Decoding Performance Bounds for LDPC Codes on Noisy Channels}

\author{\normalsize Chun-Hao Hsu and Achilleas Anastasopoulos\\
    \small Electrical Engineering and Computer Science Department\\
    \small University of Michigan, Ann Arbor, MI, 48109-2122\\
    \small email: \{chhsu, anastas\}@umich.edu
}
\date{}

\maketitle

\thispagestyle{empty}

\begin{abstract}

The asymptotic iterative decoding performances of low-density
parity-check (LDPC) codes using min-sum (MS) and sum-product (SP)
decoding algorithms on memoryless binary-input output-symmetric
(MBIOS) channels are analyzed in this paper. For MS decoding, the
analysis is done by upper bounding the bit error probability of the
root bit of a tree code by the sequence error probability of a
subcode of the tree code assuming the transmission of the all-zero
codeword. The result is a recursive upper bound on the bit error
probability after each iteration. For SP decoding, we derive a
recursively determined lower bound on the bit error probability
after each iteration. This recursive lower bound recovers the
density evolution equation of LDPC codes on the binary erasure
channel (BEC) with inequalities satisfied with equalities. A
significant implication of this result is that the performance of
LDPC codes under SP decoding on the BEC is an upper bound of the
performance on all MBIOS channels with the same uncoded bit error
probability. All results hold for the more general multi-edge type
LDPC codes.

\end{abstract}

\section{Introduction}
\label{intro}

Low-density parity-check (LDPC) codes, first introduced by
Gallager~\cite{Ga62}, have been shown to be powerful channel codes
under iterative decoding by numerous simulations in the literature.
However, the loopy graphical structure of the LDPC codes generally
prohibits their iterative decoding performance from being exactly
analyzed. This problem has been partially solved in~\cite{RiUr01} by
considering the asymptotic average performance of an ensemble of
LDPC codes when their codeword length goes to infinity. In
particular, the authors in~\cite{RiUr01} proved that the performance
of each code in the ensemble asymptotically approaches the average
performance of the ensemble with loop-free local structure within
finite iterations as the codeword length goes to infinity. This
result gave birth to the density evolution (DE) method proposed
in~\cite{RiUr01}, and was used to provide the exact asymptotic
performance of LDPC codes after an arbitrary number of iterations on
memoryless binary-input output-symmetric (MBIOS) channels.

A major drawback of the DE method is that the evolved densities in
general require an infinite dimensional description. Therefore, DE
is not suitable for the derivation of analytical performance bounds.
To solve this problem, several approaches have been proposed that
track the evolution of the densities projected to some specific
finite dimensional space. These include the Gaussian
approximation~\cite{ChRiUr01}, the extrinsic information transfer
(EXIT) chart~\cite{Br01}, and the generalized EXIT
(GEXIT)~\cite{MeMoUr05} chart methods. Unfortunately, since the EXIT
and GEXIT chart methods still require numerical calculations, and
the Gaussian approximation does not imply any upper or lower bounds
on the exact performance, these results still can not be used to
analytically bound the code performance.

In~\cite{LeTrZiCo05}, the authors propose to map the evolved
densities to Bhattacharryya parameters, which are further used to
bound the bit error probability of the LDPC codes under sum-product
(SP) decoding~\cite{Wi96} on MBIOS channels. It turns out that an
upper bound on the Bhattacharryya parameters can be obtained by a
recursion involving only one-dimensional real numbers, so the whole
result can be used to determine a guaranteed decoding capability for
the LDPC codes. In this paper, we improve this result
of~\cite{LeTrZiCo05} by showing that the same recursively determined
upper bound on the bit error probability of the LDPC codes not only
holds for the SP decoding, but also holds for the min-sum (MS)
decoding~\cite{Wi96}. This result is attained by upper bounding the
probability of error of the root bit of a tree code by a sequence
error probability of a subcode of the tree code, and then using the
union bound. Therefore, the whole proof does not involve the
Bhattacharryya parameters.

Then, we turn our attention to SP decoding and derive a recursive
lower bound on the probability of bit error after each iteration.
This recursive lower bound becomes exact and recovers the
one-dimensional density evolution equation on the binary erasure
channel (BEC). Exploiting the resemblance of our recursion to the DE
equation on the BEC, we prove that LDPC codes under SP decoding on
an MBIOS channel with uncoded bit error probability $P_0$ always
have an equal or worse performance than that on the BEC with erasure
probability $\epsilon = 2P_0$. This result should be compared with
the result in~\cite{JiRi05}, where a tighter recursive upper bound
than that in~\cite{LeTrZiCo05} on the Bhattacharryya parameters
associated with the outgoing message of bits (and hence an upper
bound on the bit error probability) after each iteration is derived,
which also exactly recovers the density evolution equation on the
BEC. Therefore, the iterative decoding analysis of LDPC codes on the
BEC can be used to bound the performance of LDPC codes from below
and above on all MBIOS channels via our result and the result
in~\cite{JiRi05}. Note also that, due the nature of the proofs of
the main lemmas, this connection between BEC and MBIOS channels is
also true for the general family of multi-edge type LDPC
codes~\cite{RiUr04}, including the irregular repeat-accumulate (IRA)
codes~\cite{JiKhMc00, PfSaUr05} and the low-density parity-check and
generator matrix (LDPC-GM) codes~\cite{HsAn05d}.

The remaining of this paper is structured as follows. In
Section~\ref{prelim}, we review the preliminary background on MBIOS
channels and the asymptotic analysis of LDPC codes. Then, we present
our asymptotic performance analysis of LDPC codes on MBIOS channels
under MS and SP decoding in Section~\ref{MSA} and~\ref{SPA},
respectively. Finally, we conclude this paper in
Section~\ref{conclusion}.

\section{Preliminaries}
\label{prelim}

If $x \in \{0, 1\}$ and $y \in \mathbb{R}$ are the input and output
symbols, respectively, of an MBIOS channel with conditional density
function $f(y|x)$, then we have the following symmetry condition:
\begin{align} \label{symy}
f(y|0) = f(-y|1), \quad \forall y \in \mathbb{R}
\end{align}
Note that all our results hold also when $y$ is a discrete random
variable, in which case, we simply treat $f(y|x)$ as a conditional
probability mass function and integration signs as summation signs
wherever necessary. However, for convenience, we will assume $y$
to be a continuous random variable throughout this paper. To
analyze the asymptotic average iterative decoding performance of a
$(\lambda, \rho)$ irregular LDPC ensemble when the codeword length
goes to infinity, where $\lambda(x) \triangleq \sum_{i=1}^\infty
\lambda_i x^{i-1}$, and $\rho(x) \triangleq \sum_{i=1}^\infty
\rho_i x^{i-1}$ are the standard variable and check node degree
distributions, respectively, from the edge perspective as defined
in~\cite{RiShUr01}, it is shown in~\cite{RiUr01} that we can as
well consider the cycle-free case. In this case, the probability
of bit error after $l$ decoding iterations is the probability of
decoding error of the root bit on the tree of $l+1$ (variable
node) levels whose construction is dictated by the degree
distributions $(\lambda, \rho)$ as in~\cite{RiShUr01}. Due to the
symmetry condition of MBIOS channels, we can also assume without
loss of generality that the all-zero codeword is transmitted.
Hence, in the following two sections, we will analyze the
asymptotic MS and SP iterative decoding performance of the
$(\lambda, \rho)$ LDPC ensemble on MBIOS channels by considering
the corresponding tree codes and assuming the transmission of the
all-zero codeword.

\section{Min-Sum Decoding Performance Analysis}
\label{MSA}

Consider an arbitrary binary tree code whose codebook is $\set{C} =
\{\ve{c}_0=\ve{0}, \ve{c}_1, \ldots, \ve{c}_M\}$, where $\ve{c}_i =
(c_{i1}, c_{i2}, \ldots, c_{in})$ is a codeword of length $n$ for
all $i$, and $\ve{c}_0$ is the all-zero codeword. Let $c_{i1}$ be
the root bit of this tree for all $i$, $\ve{x} = (x_1, x_2, \ldots,
x_n)$ the transmitted codeword, and $\ve{y} = (y_1, y_2, \ldots,
y_n)$ the received sequence from an MBIOS channel with conditional
probability mass function $p(y|x)$. When min-sum (MS) decoding is
performed on this tree, it essentially performs maximum likelihood
sequence detection (MLSqD) on the whole sequence to produce an
estimate $\hat{\ve{c}} = (\hat{c}_1, \hat{c}_2, \ldots, \hat{c}_n)$.
Therefore, if we define the decision region for the codeword
$\ve{c}_i$ as
\begin{align}
\set{U}_i \triangleq \bigcap_{k\neq i} \set{U}_{ik},
\end{align}
where
\begin{align} \label{pairwise_U}
\set{U}_{ik} \triangleq \left\{\ve{y} \in \mathbb{R}^n | \prod_{j =
1}^n p(y_j|c_{ij}) \geq \prod_{j = 1}^n p(y_j|c_{kj})\right\},
\end{align}
then the probability of the root bit being in error under MS
decoding assuming $\ve{x} = \ve{0}$ is
\begin{equation}
P_b^{MS} = Pr(\hat{c}_1=1|\ve{x} = \ve{c}_0) = Pr\left(\ve{y} \in
\bigcup_{i,c_{i1}=1} \set{U}_{i}|\ve{x} = \ve{c}_0\right).
\end{equation}
We would like to find a compact representation of the set
\begin{align}
\set{U} \triangleq \bigcup_{i,c_{i1}=1} \set{U}_{i},
\end{align}
or at least a superset of $\set{U}$ so that we can bound $P_b^{MS}$
from above. Note that such an upper bound is also a valid upper
bound on the probability of root bit error $P_b^{SP}$ when the SP
decoding is performed on the tree since the SP decoding essentially
performs optimal maximum likelihood decoding for each bit.

\begin{definition}
Define the subcode $\set{C}_r$ of $\set{C}$ as the set of
codewords in $\set{C}$ such that
 \begin{itemize}
 \item[(i)]  the root bit is 1, and
 \item[(ii)]  each check node with parent variable node equal to 1
              has exactly one child variable node equal to 1, and
 \item[(iii)] each check node with parent variable node equal to 0
              has all children variable nodes equal to 0.
 \end{itemize}
\end{definition}

We show how we can use this reduced codebook $\set{C}_r$ to
characterize $\set{U}$ in the following lemma.
\begin{lemma}
$\set{U} \subset \bigcup_{i, \ve{c}_i \in \set{C}_r} \set{U}_{i0}$.
\end{lemma}
\begin{proof}
Given any $\ve{y} \in \set{U} = \bigcup_{i,c_{i1}=1} \set{U}_{i}$,
there exists a $k$ such that $\ve{y} \in \set{U}_{k}$ and $c_{k 1}
= 1$. To proceed with our proof, we first carry out the following
labeling procedure on the codeword $\ve{c}_k$.
\begin{enumerate}
\item At the initial state, the root bit is labelled as ``survivor'',
and all the other variable nodes are unlabelled. Note that in this
labeling procedure, the ``survivor'' variable nodes will always
have value 1 in $\ve{c}_k$. We first consider the check nodes at
the topmost level of the tree.

\item For every check node $c$ at this level whose parent node is
labelled as ``survivor'', it must have at least one child variable
node with value 1 in $\ve{c}_k$ since ``survivor'' nodes always
have value 1 in $\ve{c}_k$. Choose an arbitrary child variable
node of $c$ with value 1 in $\ve{c}_k$, and label it as
``survivor''. Then, label the \emph{subtrees} emanating from the
other unlabelled child variable nodes of $c$ as ``dropped''.

\item If there are no check nodes at the next lower level of the tree, stop.
Otherwise, move to the check nodes at the next lower level and go
back to 2.
\end{enumerate}
As we can see after this labeling procedure, the check nodes with a
``survivor'' parent node all have exactly one ``survivor'' child
node, and the ones with a ``dropped'' parent node have purely
``dropped'' child nodes. Therefore, if we let
\begin{align}
\ve{c}_m =
\begin{cases}
\ve{0}& \mbox{for all ``dropped'' bits} \\
\ve{1} = \ve{c}_k& \mbox{for all ``survivor'' bits}
\end{cases}
\end{align}
then we have $\ve{c}_m \in \set{C}_r \subset \set{C}$. In the
following, we would like to prove that $\ve{y} \in \set{U}_{m0}$,
which completes the proof. Let
\begin{align}
\ve{c}_l =
\begin{cases}
\ve{c}_k& \mbox{for all ``dropped'' bits} \\
\ve{0}& \mbox{for all ``survivor'' bits}
\end{cases}
\end{align}
\ie, let $\ve{c}_l$ be the bitwise XOR of $\ve{c}_k$ and $\ve{c}_m$.
Then, since $\ve{c}_k$ and $\ve{c}_m$ are valid codewords in
$\set{C}$, so is $\ve{c}_l$. Moreover, since $\ve{y} \in
\set{U}_{k}$ implies $\ve{y} \in \set{U}_{kl}$, we have
from~\eqref{pairwise_U} that
\begin{align}
\prod_{j = 1}^n p(y_j|c_{kj}) \geq \prod_{j = 1}^n p(y_j|c_{lj})
\Rightarrow& \prod_{\text{all ``survivor'' bits $j$}} p(y_j|c_{kj})
\geq \prod_{\text{all ``survivor'' bits $j$}}
p(y_j|c_{lj}) \nonumber \\
\Rightarrow& \prod_{\text{all ``survivor'' bits $j$}} p(y_j|c_{mj})
\geq \prod_{\text{all ``survivor'' bits $j$}}
p(y_j|0) \nonumber \\
\Rightarrow& \prod_{j = 1}^n p(y_j|c_{mj}) \geq \prod_{j = 1}^n
p(y_j|0)
\end{align}
which proves that $\ve{y} \in \set{U}_{m0}$ as desired.
\end{proof}
This lemma shows that we can upper bound $P_b^{MS}$ by the
probability of MLSqD error $P_s^{MLSqD}$ on the reduced codebook
$\set{C}_r$ assuming that the all-zero codeword is transmitted. One
way to proceed from here is to use union bound and the fact
(see~\cite[Theorem 7.5]{Mc02} for a proof) that
\begin{align}
Pr(\ve{y} \in \set{U}_{i0}| \ve{x} = \ve{c}_0)\leq D^{w(\ve{c}_i)},
\quad \forall i
\end{align}
where $w(\ve{c}_i)$ denotes the Hamming weight of $\ve{c}_i$, and
$D$ is the Bhattacharryya parameter associated with the MBIOS
channel $p(y|x)$, to further upper bound $P_s^{MLSqD}$. For this
purpose, we would like to introduce the weight enumerator $N_l(x)$
of the reduced codebook $\set{C}_l$ of the tree code $G_l$ of
level $l+1$ associated with a randomly drawn code from the
$(\lambda, \rho)$ LDPC ensemble. Let $A_i$ be the number of
codewords of weight $i$ in $\set{C}_l$. We define $N_l(x)$ by
$N_l(x) \triangleq \sum_{i = 1}^\infty A_i x^i$. Moreover, let
$\overline{N_l(x)}$ denote the expected value of $N_l(x)$ averaged
over the whole $(\lambda, \rho)$ LDPC ensemble. We have the
following lemma.
\begin{lemma} \label{ms_main}
$\overline{N_0(x)} = x$ and
\begin{align}
\overline{N_l(x)} = \lambda(\rho'(1)\overline{N_{l-1}(x)}), \quad
\forall l \geq 1
\end{align}
where $\rho'(x)$ denotes the derivative of $\rho(x)$.
\end{lemma}
\begin{proof}
It is obvious that $\overline{N_0(x)} = x$. To prove the recursion
for $\overline{N_l(x)}$, first consider the subtree emanating from
the $i$th check node $c_i$ immediately below the root bit. Let
$Z_l^{(i)}(x)$ denote the weight enumerator of the reduced
codebook of this subtree. Since, the root bit is 1 for all
codewords in the reduced codebook $\set{C}_l$, there is exactly
one child subtree with root 1 emanating from $c_i$ for all
codewords in $\set{C}_l$. Therefore, if $c_i$ has degree $d_c$,
then we have
\begin{align} \label{ms_recur_ch}
Z_l^{(i)}(x) = (d_c - 1) N_{l-1}(x) \Rightarrow
\overline{Z_l^{(i)}(x)} = \sum_{i = 1}^\infty (i - 1) N_{l-1}(x)
\rho_i = \rho'(1) \overline{N_{l-1}(x)}.
\end{align}
Similarly, if the root bit has degree $d_v$, then we have
\begin{align} \label{ms_recur_var}
N_l(x) = \prod_{i=1}^{d_c-1} Z_l^{(i)}(x) \Rightarrow
\overline{N_l(x)} = \sum_{j=1}^\infty \overline{\prod_{i=1}^{d_c-1}
Z_l^{(i)}(x)} \lambda_j = \sum_{i=j}^\infty \prod_{i=1}^{d_c-1}
\overline{Z_l^{(i)}(x)} \lambda_j = \lambda(\overline{Z_l^{(1)}(x)})
\end{align}
where the second equality follows from the fact that the subtrees
emanating from different $c_i$'s are generated independently, and
the third equality follows from the fact that
$\overline{Z_l^{(i)}(x)}$ does not depend on $i$ as shown
in~\eqref{ms_recur_ch}. Combining~\eqref{ms_recur_ch}
and~\eqref{ms_recur_var}, the lemma is proved.
\end{proof}
Using this lemma and the union bound on $P_s^{MLSqD}$, we have the
following theorem.
\begin{theorem} \label{main1}
Given any $(\lambda, \rho)$ LDPC ensemble, let $P_l^{MS}$ and
$P_l^{SP}$ be its asymptotic (as the codeword length approaches
infinity) average bit error probability after $l$ iterations under
MS and SP decoding, respectively, on an MBIOS channel with
Bhattacharryya parameter $D$. If we define the sequence $\{z_l\}_{l
= 0}^\infty$ by $z_0 = D$, and $z_l = \lambda(\rho'(1)z_{l-1})$, for
all $l \geq 1$, then we have $P_l^{SP} \leq P_l^{MS} \leq z_l$, for
all $l \geq 0$.
\end{theorem}
\begin{proof}
From Lemma~\ref{ms_main} we see that $z_l = \overline{N_l(D)}$ for
all $l$. Now, the lemma follows from the union bound on
$P_s^{MLSqD}$ as discussed above.
\end{proof}
A similar result to Theorem~\ref{main1} is established
in~\cite[Lemma 1]{LeTrZiCo05}. However, Theorem~\ref{main1} differs
from~\cite[Lemma 1]{LeTrZiCo05} in two aspects. First,
Theorem~\ref{main1} holds for both MS and SP decoding
while~\cite[Lemma 1]{LeTrZiCo05} holds only for the SP decoding.
Second, we did not keep track of the evolution of the Bhattacharryya
parameters, which are used in~\cite{LeTrZiCo05} to bound $P_l^{SP}$
for all $l$.

\section{Sum-Product Decoding Performance Analysis}
\label{SPA}

Let
\begin{align}
m = \log \frac{f(y|0)}{f(y|1)}
\end{align}
be the log-likelihood ratio (LLR) of the input variable $x$ given
the output variable $y$ of the MBIOS channel $f(y|x)$. Moreover, let
$M$ be the random variable whose realization is $m$ assuming $x =
0$, and $g(m)$ be the probability density function (pdf) of $M$.
Then the symmetry condition~\eqref{symy} becomes
\begin{align} \label{symm}
g(-m) = e^{-m}g(m), \quad \forall m \in \mathbb{R}
\end{align}
Define the probability of error $P_e(M)$ under ML decoding of the
LLR $M$ as follows
\begin{align}
P_e(M) = \int_{-\infty}^{0} g(m) dm.
\end{align}
We have the following lemma.
\begin{lemma} \label{basic}
\begin{align}
E\left[\tanh\frac{|M|}{2}\right] = 1 - 2P_e(M)
\end{align}
\end{lemma}
\begin{proof}
\begin{align}
E\left[\tanh\frac{|M|}{2}\right] =& \int_0^{\infty}
\left(\tanh\frac{m}{2}\right)\left[g(m) + g(-m)\right] dm \nonumber
\\
=& \int_0^{\infty} \left(\frac{1 - e^{-m}}{1 +
e^{-m}}\right)\left[g(m) + e^{-m}g(m)\right] dm \nonumber \\
=& \int_0^{\infty} \left(1 - e^{-m}\right) g(m) dm \nonumber \\
=& \int_0^{\infty} g(m) dm - \int_{-\infty}^0 g(m) dm \nonumber \\
=& 1 - 2P_e(M)
\end{align}
\end{proof}

Now, consider the SP decoding on a tree code used on an MBIOS
channel. As shown in~\cite{RiUr01}, all the SP decoding messages can
be represented by the LLR's and satisfy the symmetry
condition~\eqref{symm}. Assuming that the all-zero codeword is
transmitted, we have the following lemma describing the relationship
between the probability of ML decoding errors associated with the
incoming and outgoing messages of a check node. Note that in the
following, we will use capital letters to denote random variables,
whose realizations are denoted by the corresponding lower-case
letters.
\begin{lemma} \label{check_lemma}
Let $c$ be a check node of degree $d_c$. Furthermore, let $M_c$ be
the outgoing message on an edge, and $M_1, M_2, \ldots, M_{d_c-1}$
the incoming messages from the other edges. Assuming all the
incoming messages are independent with each other, we have
\begin{align} \label{check_eq}
1 - 2P_e(M_c) = \prod_{i=0}^{d_c-1} \left[1 - 2P_e(M_i)\right]
\end{align}
\end{lemma}
\begin{proof}
It follows from Lemma~\ref{basic}, the fact that under SP
decoding~\cite{RiUr01},
\begin{align}
\tanh\frac{|m_c|}{2} = \prod_{i = 1}^{d_c - 1} \tanh\frac{|m_i|}{2}
\end{align}
and the independence of the incoming messages.
\end{proof}
A similar relationship for a variable node is also derived. However,
since the derivation is more complicated than the one for a check
node, we first consider the simple case where the number of incoming
messages is 2.
\begin{lemma} \label{var_lemma}
Let $v$ be a variable node. Furthermore, let $M_v$ be the outgoing
message on an edge, and $M_1$ and $M_2$ the incoming messages from
either the other edges or the channel. Assuming $M_1$ and $M_2$ are
the only incoming messages and independent with each other, we have
\begin{align} \label{var_eq}
2P_e(M_v) \geq \prod_{i = 1}^2 \left[2P_e(M_i)\right]
\end{align}
\end{lemma}
\begin{proof}
From Lemma~\ref{basic}, we have
\begin{align} \label{check_eq1}
&2P_e(M_v) \geq \prod_{i = 1}^2 \left[2P_e(M_i)\right] \nonumber \\
\Leftrightarrow& 1 - E\left[\tanh\frac{|M_v|}{2}\right] \geq \left(1
- E\left[\tanh\frac{|M_1|}{2}\right]\right)\left(1 -
E\left[\tanh\frac{|M_2|}{2}\right]\right) \nonumber \\
\Leftrightarrow& E\left[\tanh\frac{|M_v|}{2}\right] \leq
E\left[\tanh\frac{|M_1|}{2}\right] +
E\left[\tanh\frac{|M_2|}{2}\right] -
E\left[\tanh\frac{|M_1|}{2}\right]E\left[\tanh\frac{|M_2|}{2}\right]
\end{align}
Since under SP decoding, we have from~\cite{RiUr01} that $m_v = m_1
+ m_2$, the left hand side of~\eqref{check_eq1} becomes
\begin{align}
&\int_{m_1 + m_2 \geq 0} \left[1 - e^{-(m_1 +
m_2)}\right]g_1(m_1)g_2(m_2) dm_1 dm_2 \nonumber \\
=& \int_{m_1\geq 0, m_2 \geq 0} \left[1 - e^{-(m_1 +
m_2)}\right]g_1(m_1)g_2(m_2) dm_1 dm_2 \nonumber \\
&+ \int_{m_1\geq -m_2, m_2 < 0} \left[1 - e^{-(m_1 +
m_2)}\right]g_1(m_1)g_2(m_2) dm_1 dm_2 \nonumber \\
&+ \int_{m_2\geq -m_1, m_1 < 0} \left[1 - e^{-(m_1 +
m_2)}\right]g_1(m_1)g_2(m_2) dm_1 dm_2
\end{align}
where $g_1$ and $g_2$ are the pdf's of $M_1$ and $M_2$,
respectively. On the other hand, the right hand side becomes
\begin{align}
&\int_{m_1 \geq 0} \left(1 - e^{-m_1}\right)g_1(m_1) dm_1 +
\int_{m_2 \geq 0} \left(1 - e^{-m_2}\right)g_2(m_2) dm_2 \nonumber
\\
&- \int_{m_1 \geq 0, m_2 \geq 0} \left(1 - e^{-m_1}\right)\left(1 -
e^{-m_2}\right)g_1(m_1)g_2(m_2) dm_1dm_2 \nonumber \\
=& \int_{m_1 \geq 0, m_2 \geq 0} \left[1 - e^{-(m_1+m_2)}\right]
g_1(m_1) g_2(m_2) dm_1 dm_2 \nonumber \\
&+ \int_{m_1 \geq 0, m_2 < 0} \left(1 -
e^{-m_1}\right)g_1(m_1)g_2(m_2) dm_1 dm_2 + \int_{m_2 \geq 0, m_1 <
0} \left(1 - e^{-m_2}\right)g_1(m_1)g_2(m_2) dm_1 dm_2
\end{align}
Now, since
\begin{align}
\int_{m_i \geq 0, m_j < 0} \left(1 - e^{-m_i}\right)g_i(m_i)g_j(m_j)
dm_i dm_j \geq& \int_{m_i \geq -m_j, m_j < 0} \left(1 -
e^{-m_i}\right)g_i(m_i)g_j(m_j) dm_i dm_j \nonumber \\
\geq& \int_{m_i \geq -m_j, m_j < 0} \left(1 -
e^{-m_i}e^{-m_j}\right)g_i(m_i)g_j(m_j) dm_i dm_j
\end{align}
for $(i, j)$ equals $(1, 2)$ and $(2, 1)$, \eqref{check_eq1} is
true, and the lemma is proved.
\end{proof}
Armed with the previous lemma, we can proceed to the general case
where the variable node has degree $d_v$.
\begin{corollary} \label{var_coro}
Let $v$ be a variable node of degree $d_v$. Furthermore, let $M_0$
be the incoming message from the channel, $M_v$ the outgoing message
on an edge, and $M_1, M_2, \ldots, M_{d_v-1}$ the incoming messages
from the other edges. Assuming all the incoming messages are
independent with each other, we have
\begin{align}
2P_e(M_v) \geq \prod_{i = 0}^{d_v-1} \left[2P_e(M_i)\right]
\end{align}
\end{corollary}
\begin{proof}
It follows directly from Lemma~\ref{var_lemma} and induction.
\end{proof}
For a tree code used on an MBIOS channel under SP decoding, all the
incoming messages to the variable and check nodes are independent
with each other. Hence, Lemma~\ref{check_lemma} and
Corollary~\ref{var_coro} can be used to characterize the evolution
of the probability of error associated with the messages after
processes of the variable and check nodes on a tree, and imply the
following theorem.
\begin{theorem}
For the $(\lambda, \rho)$ LDPC ensemble used on an MBIOS channel,
the probability of bit error $P_l$ associated with the outgoing
message of any variable node after the $l$th decoding iteration
asymptotically satisfies
\begin{align} \label{main2}
2P_l \geq 2P_0 \lambda\left(1 - \rho\left(1 - 2P_{l-1}\right)\right)
\end{align}
where $P_0$ is the uncoded bit error probability under ML decoding
of the channel.
\end{theorem}
\begin{proof}
As discussed in Section~\ref{prelim}, the probability of error
associated with the outgoing message of any variable node after the
$l$th decoding iteration is, asymptotically as the codeword length
goes to infinity, the one of the root variable node of a tree of
level $l+1$. Since all the variable and check nodes in the tree have
exactly the same degree distributions $\lambda$ and $\rho$,
respectively, and the channel is memoryless, all the incoming
messages from child nodes to a parent node are independent and
identically distributed. Hence, if we let $c$ be a check node in the
tree, $M_c$ its outgoing message, and $M_v$ one of its incoming
message, then we have from Lemma~\ref{check_lemma} that
\begin{align} \label{main2_check}
2P_e(M_c) = 1 - \sum_{i=1}^\infty \left[1 - 2P_e(M_v)\right]^{(i-1)}
\rho_i = 1 - \rho(1 - 2P_e(M_v))
\end{align}
Similarly, if we let $v$ be a variable node in the tree, $M_c$ its
outgoing message, $M_0$ the incoming message from the channel, and
$M_v$ one of its incoming messages from its child nodes, then we
have from Corollary~\ref{var_coro} that
\begin{align} \label{main2_var}
2P_e(M_v) \geq 2P_e(M_0)\sum_{i=1}^\infty
\left[2P_e(M_c)\right]^{(i-1)} \lambda_i = 2P_e(M_0)\lambda(1 -
2P_e(M_c))
\end{align}
Combining~\eqref{main2_check}, ~\eqref{main2_var}, and the fact that
$\rho$ is a monotonically increasing function (since all $\rho_i$'s
are nonnegative), the theorem is proved.
\end{proof}
Notice that on the binary erasure channel (BEC) of erasure
probability $\epsilon$, the probability of uncoded bit error of this
channel is $\epsilon/2$. Hence, \eqref{main2} is satisfied with
equality, and recovers the well-known DE equation
\begin{align} \label{old_de}
x_l = \epsilon \lambda\left(1 - \rho\left(1 - x_{l-1}\right)\right),
\quad x_0 = \epsilon
\end{align}
where $x_l$ is the bit erasure probability after the $l$th iteration
on the BEC. Using this fact, we have the following corollary.
\begin{corollary} \label{main2_imp}
For any $(\lambda, \rho)$ LDPC ensemble, its asymptotic SP decoding
performance on the MBIOS channel with uncoded bit error probability
$P_0$ is always worse than its asymptotic SP decoding performance on
the BEC with erasure probability $\epsilon \leq 2P_0$.
\end{corollary}
\begin{proof}
From~\eqref{main2} and~\eqref{old_de}, we have $x_l \leq 2P_l$ for
all $l \geq 0$ when $\epsilon \leq 2P_0$. Since the erasure
probability is two times the bit error probability, the corollary is
proved.
\end{proof}
Notice that, Lemma~\ref{check_lemma} and Corollary~\ref{var_coro}
can also be used in the more general family of multi-edge type LDPC
codes~\cite{RiUr04}, including the irregular repeat-accumulate (IRA)
codes~\cite{JiKhMc00, PfSaUr05} and the low-density parity-check and
generator matrix (LDPC-GM) codes~\cite{HsAn05d}, to produce similar
results. Hence, Corollary~\ref{main2_imp} is not restricted to the
irregular LDPC codes, but also holds for the general multi-edge type
LDPC codes.

\section{Conclusion}
\label{conclusion}

In this paper, we analyze the asymptotic performance of LDPC codes
under MS and SP decoding on MBIOS channels. This is done by upper
bounding the bit error probability of the root bit of the tree code
associated with the $(\lambda, \rho)$ LDPC ensemble assuming the
all-zero codeword is transmitted. When MS decoding is performed on
this tree code, we upper bound the probability of the root bit being
in error by the probability of sequence error under ML decoding of a
subcode of the tree code. A recursive equation describing the
evolution of the weight enumerator of this subcode after each
iteration is then derived and used in a union bound to bound the ML
decoded sequence error of this subcode. As a result, we obtain a
recursive upper bound on the bit error probability as a function of
the number of iterations for the LDPC codes under MS decoding on
MBIOS channels. Note that these upper bounds are also upper bounds
for the SP decoding since SP decoding is optimal on the bit error
probability for tree codes. This result is very similar
to~\cite[Lemma 1]{LeTrZiCo05} with the difference being that we
establish it not only for the SP decoding, but also for the MS
decoding, and that we obtain it via a totally different approach.

When SP decoding performance is considered, we derive a recursive
lower bound on the probability of bit error as a function of the
number of iterations. Note that this is also a lower bound for the
MS decoding due to the optimality of SP decoding on the bit error
probability. More significantly, this recursion recovers the DE
equation on the BEC for LDPC codes with the lower bound being an
exact equality. This further implies that the SP decoding
performance of LDPC codes on the BEC can serve as a upper bound of
the performance on all MBIOS channels with the same probability of
uncoded bit error. This result is also true for the more general
multi-edge type LDPC codes, including IRA and LDPC-GM codes, since
the main ingredient in the proof, i.e., Lemma~\ref{check_lemma} and
Corollary~\ref{var_coro}, can also be utilized for these codes.

\bibliographystyle{IEEEbib}
\bibliography{hsuref}

\end{document}